\documentclass[aps,pra,twocolumn,reprint]{revtex4-1}
\usepackage[usenames]{color}
\DeclareMathAlphabet{\mathpzc}{OT1}{pzc}{m}{it}
\newcommand{\comment}[1]{}
\usepackage{graphicx}
\usepackage{epsfig}
\usepackage{amsmath}
\usepackage{mathrsfs}
\usepackage{amsbsy}
\usepackage{color}
\usepackage{eucal}
\usepackage{amsfonts}
\newcommand{\bra}[1]{\langle {#1} |}
\newcommand{\ket}[1]{| {#1} \rangle}

\newcommand{\ketn}[1]{ {#1} \rangle}
\newcommand{\br}{{\bf r}}
\newcommand{\bp}{{\bf p}}
\newcommand{\bs}{{\boldsymbol \sigma}}
\newcommand{\bsp}{{\boldsymbol \sigma_\perp}}
\newcommand{\bsx}{{\boldsymbol \sigma_x}}
\newcommand{\bsy}{{\boldsymbol \sigma_y}}
\newcommand{\bsz}{{\boldsymbol \sigma_z}}
\newcommand{\bmu}{{\boldsymbol \chi}}
\newcommand{\bk}{{\bf k}}
\newcommand{\bv}{{\bf v}}
\newcommand{\bpsi}{{\boldsymbol \Psi}}
\newcommand{\bpsil}{{\boldsymbol \psi}}
\newcommand{\bphi}{{\boldsymbol \Phi}}
\newcommand{\bet}{{\boldsymbol \eta}}
\newcommand{\id}{{\boldsymbol 1}}

\newcommand{\ce}{{\cal E}}
\newcommand{\cf}{{ F}}

\begin{document}
\title{Order by Disorder in Spin-Orbit Coupled Bose-Einstein Condensates}
\date{\today}

\author{Ryan Barnett,$^1$ Stephen Powell,$^1$ Tobias Gra\ss,$^2$
  Maciej Lewenstein,$^{2,3}$, and S. Das Sarma$^1$}
\affiliation{$^1$Joint Quantum
Institute and Condensed Matter Theory Center,
Department of Physics, University of Maryland, College
Park, Maryland 20742-4111, USA}
\affiliation{$^2$ICFO-Institut de Ci\`encies
Fot\`oniques, Mediterranean
Technology Park, 08860 Castelldefels (Barcelona), Spain}
\affiliation{$^3$ICREA-Instituci\'o Catalana de Recerca i Estudis Avan\c
cats, Lluis Companys 23, 08010 Barcelona, Spain}

\begin{abstract}
Motivated by recent experiments, we investigate the system of
isotropically-interacting bosons with Rashba spin-orbit coupling.  At
the non-interacting level, there is a macroscopic ground-state
degeneracy due to the many ways bosons can occupy the
Rashba spectrum.  Interactions treated at the mean-field level
restrict the possible ground-state configurations, but there remains
an accidental degeneracy not corresponding to any symmetry of the
Hamiltonian, indicating the importance of fluctuations.  By finding
analytical expressions for the collective excitations in the
long-wavelength limit and through numerical solution of the full
Bogoliubov- de Gennes equations, we show that the system condenses into
a single momentum state of the Rashba spectrum via the mechanism of
order by disorder.  We show that in 3D the quantum depletion for this
system is small, while the thermal depletion has an infrared
logarithmic divergence, which is removed for finite-size systems. In
2D, on the other hand, thermal fluctuations destabilize the system.
\end{abstract}
\maketitle

\section{Introduction and Overview}

Multicomponent condensates of ultracold atoms offer rich physical
systems due to the interplay between superfluidity and internal
degrees of freedom \cite{Lewenstein07}.  Recently, through the use of
synthetic gauge fields, two-component bosons with spin-orbit (SO)
coupling have been engineered in the ultracold laboratory
\cite{lin11}.  SO coupling in solid-state materials has a long history
and is responsible for a variety of interesting physical effects, with notable
examples including the spin Hall effect \cite{kato04} and topological
insulators \cite{hasan10}.  In addition, SO-coupled materials have diverse applications including spintronics
\cite{zutic04}.  The newer bosonic counterpart of SO-coupled systems
using ultracold atoms have no analog in solid-state systems and are thus
expected to exhibit genuinely new physics.  SO-coupled cold atomic
systems have also received considerable recent theoretical attention 
\cite{wu08,zhang08,stanescu08,ho10,larson10,
  merkl10,wang10,yip11,ozawa11, sau11,jiang11,,van11,zhang11,gopalakrishnan11,hu11,sinha11},
investigating topics such as spin-striped states \cite{ho10, wang10,
yip11}, fragmentation \cite{stanescu08,gopalakrishnan11}, and
the realization of Majorana fermions \cite{zhang08,sau11, jiang11}.

\begin{figure}
\includegraphics[width=2in]{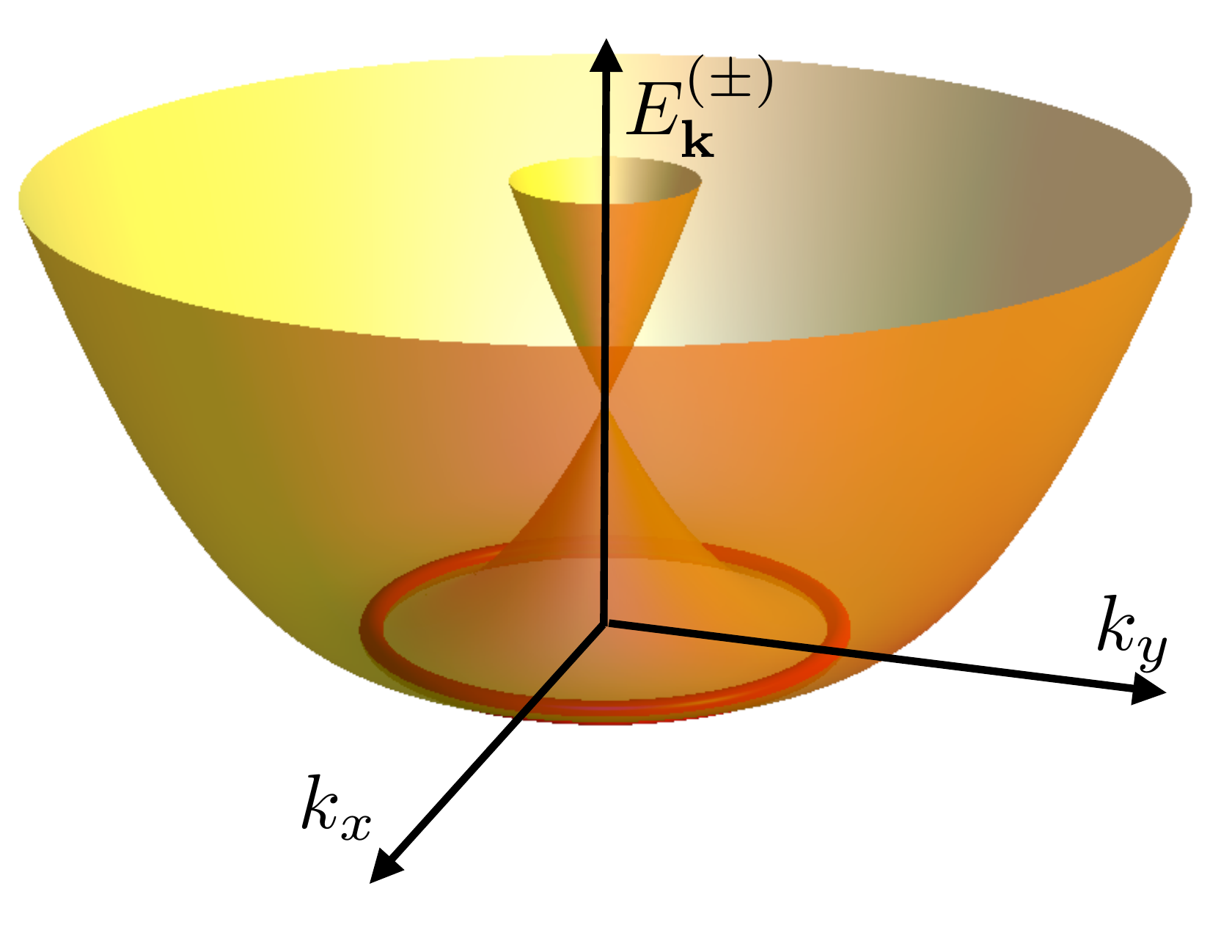}
\caption{(Color online) The non-interacting (Rashba) energy spectrum
  of the Hamiltonian Eq.~(\ref{Eq:H0}) with $k_z=0$.  The red circle
  indicates the degenerate lowest-energy single-particle states.}
\label{Fig:nispec}
\end{figure}

Recent experiments \cite{lin11} have realized a special combination
of Rashba \cite{bychkov84} and Dresselhaus \cite{dresselhaus55} SO
coupling in ultracold atoms.
There are also promising proposals to realize more general
non-Abelian gauge fields like pure Rashba (c.f. \cite{dalibard10}), or
even SU$(N)$ gauge fields that provide a toolbox for topological
insulators \cite{Mazza11}.  The conceptually simple system of a Rashba
SO-coupled Bose-Einstein Condensate with isotropic interactions (RBEC)
has surprisingly rich physics. The non-interacting system has a
macroscopic ground state degeneracy as shown in Fig.~\ref{Fig:nispec}.
Interactions at the mean-field level partially remove this degeneracy,
but there remains an `accidental' degeneracy not corresponding to any
underlying symmetry of the system.  Specifically, mean-field theory
predicts a superposition of condensates of opposite momenta with their
relative amplitudes and phases unspecified.

In this work we show how fluctuations remove this accidental degeneracy and
select a unique ground state  (up to overall symmetries) through the mechanism of `order by
disorder' \cite{villain80}. Although the phenomenon of order by
disorder has been theoretically accepted and discussed within the
context of classical spin models \cite{villain80,moessner98}, quantum
magnetism \cite{henley89,bergman07} and ultracold atoms
\cite{turner07,song07,zhao08,toth10}, experimental 
demonstrations are, at best, scarce 
\cite{chalker11}.
In contrast to the original proposal
\cite{villain80}, the degeneracy lifting we find is primarily quantum
driven.  We determine the fluctuation spectrum by numerically solving
the coupled Bogoliubov-de Gennes equations.  The resulting modes are
integrated over to obtain the free energy as a function of the
relative condensate weights and temperature.  With this we show that
fluctuations select a state with all bosons condensing into a single
momentum state in the Rashba spectrum.  We estimate the energy splitting per
particle due to fluctuations for typical experimental parameters to be
on the order of 100 pK.  While this splitting is smaller than typical
condensate temperatures, it is the total energy which determines the
ground state, so this effect should be readily observable provided the
RBEC model can be realized.

\section{Definition of Hamiltonian and Mean-Field Ground States}

The Hamiltonian  describing non-interacting bosons in 3D with SO
coupling reads
\begin{equation}
\label{Eq:H0} {\cal H}_0=\int d\br \hat{\bpsi}^\dagger(\br)
\left(\frac{p^2}{2m} - \frac{Q}{ m} \bsp \cdot \bp \right)
\hat{\bpsi}(\br),
\end{equation}
where
$\hat{\bpsi}(\br)=(\Psi_\uparrow(\br),\Psi_\downarrow(\br))^T$ is
a two-component bosonic field operator, ${\bf p}$ is the momentum
operator, $Q$ is the magnitude of the SO coupling, and $\bsp$ is a
vector composed of Pauli matrices as $\bsp=(\bsx,\bsy,0)$ (we set
$\hbar=1$).  The SO coupling in Eq.~(\ref{Eq:H0}) is equivalent to
the Rashba form \cite{bychkov84} through a $90^\circ$ spin rotation.
The single-particle eigenstates of Eq.~(\ref{Eq:H0}) have spins
pointing either parallel or antiparallel to their momenta in the
$xy$ plane, and up to a constant have energies $
E_{\bk}^{(\pm)}=\frac{1}{2m} \left( (k_\perp \pm Q )^2 +
k_z^2\right) $, where $\bk_\perp = (k_x,k_y,0)$. Clearly, there is
a ring in momentum space of degenerate lowest-energy states with
$k_z=0$ and $|\bk_\perp|=Q$ (Fig.~\ref{Fig:nispec}).
Correspondingly, there is a macroscopic number of ways $N$
non-interacting bosons can occupy this manifold of states.

For the interacting portion of the Hamiltonian we take the simplest
SU(2) invariant form
\begin{equation}
{\cal H}_{\rm int}= \int d \br \left(\frac{g}{2}\left[
\hat{\rho}(\br) \right]^2 -\mu \hat{\rho}(\br) \right),
\end{equation}
where $\hat{\rho}(\br) = \hat{\bpsi}^\dagger(\br)
\hat{\bpsi}(\br)$, $\mu$ is the chemical potential, and
$g=\frac{4\pi a}{m}$ where $a$ is an effective scattering length.
At the mean-field level one replaces the operators by c-numbers
$\hat{\bpsi}(\br) \rightarrow \bpsi (\br)$. The  states that
minimize the kinetic energy, Eq.~(\ref{Eq:H0}), are in general given
by
\begin{equation}
\label{Eq:gen1}
\bpsi(\br) = \sum_{|\bk_\perp|=Q, k_z=0} A_{\bk} \frac{e^{i\bk \cdot \br}}{\sqrt{2}}
\left(
\begin{array}{c}
1 \\ e^{i \varphi_{\bk}}
\end{array}
\right)
\end{equation}
where $A_{\bk}$ are arbitrary coefficients and
$\tan(\varphi_{\bk})=k_x/k_y$.  Minimizing the interaction energy
restricts the mean-field states of Eq.~(\ref{Eq:gen1}) to have a
constant density, $\rho(\br) = \bpsi^\dagger(\br) \bpsi(\br)\equiv
\rho_0$.   Placing this constraint on states in
Eq.~(\ref{Eq:gen1}), one finds that $\bpsi(\br)$ can have at
most two nonzero coefficients $A_\bk$ occurring at opposite momenta. 
This can be shown by setting each non-zero wavevector component of $\bpsi^\dagger(\br) \bpsi(\br)$
to zero.
Without loss of generality, we take the momenta
to point along the $x$-axis and thereby obtain the state
\begin{equation}
\label{Eq:gen2}
\bpsi(\br)= \sqrt{\frac{\rho_0}{2}} \left( a  e^{i Q x}
\left(
\begin{array}{c}
1 \\
1
\end{array}
\right)
+
be^{-i Q x}
\left(
\begin{array}{c}
-1 \\
1
\end{array}
\right) \right),
\end{equation}
where $|a|^2+|b|^2=1$.  We can take $a$ and $b$ to be real and
parametrized as $a=\cos\left(\frac{\theta}{2}\right)$ and
$b=\sin\left(\frac{\theta}{2}\right)$ since changing the phases of
$a$ and $b$ amounts to position displacements and overall phase
shifts of $\bpsi(\br)$ in Eq.~(\ref{Eq:gen2}).  The selection of $(a,b)$ 
as a result of spin-symmetry breaking interactions (which is resolved
at the mean-field level) was worked out in \cite{wang10}.  In
contrast, in this work there remains a degeneracy at the mean-field level.

\section{Calculation of Collective Excitations}

The degeneracy with respect to  $\theta$ is accidental, i.e.   it
does not correspond to any symmetry of the Hamiltonian $ {\cal H}
= {\cal H}_0 + {\cal H}_{\rm int}. $ We thus expect quantum
fluctuations about the mean-field state Eq.~(\ref{Eq:gen2}) to
remove this degeneracy and to select a unique ground state through
the order-by-disorder mechanism. To this end, we write 
$\hat{\bpsi}(\br) = \bpsi(\br) + \hat{\bpsil}(\br) $ and perform a
Bogoliubov expansion of ${\cal H}$ to quadratic order in
$\hat{\bpsil}(\br)$.  Up to a constant the interaction Hamiltonian
becomes ${\cal H}_{\rm int}=\frac{g}{2} \int d \br \left[ \delta
  \hat{\rho}(\br)\right]^2$
where
$
\delta \hat{\rho}(\br) = \bpsi^\dagger(\br) \hat{\bpsil}(\br) +
\hat{\bpsil}^\dagger(\br) \bpsi(\br).
$
It proves useful to transform to
the variables $\hat{\bmu}(\br)=(\hat{\chi}_\uparrow(\br),\hat{\chi}_\downarrow(\br))^T$
where
$
\hat{\bmu}(\br)=e^{i\bsy \frac{\theta}{2}} e^{-i \bsz Qx} e^{i \bsy
  \frac{\pi}{4}} \hat{\bpsil}(\br)
$ for which the interaction Hamiltonian takes the simple form $
{\cal H}_{\rm int} = \frac{g}{2} \int d\br \left(
  \hat{\chi}_{\uparrow}(\br)+ \hat{\chi}_{ \uparrow}^\dagger(\br)\right)^2.
$

The full Bogoliubov Hamiltonian, ${\cal H}_{\rm Bog}$,  can be written compactly in matrix form
if we introduce the four-component vector
\begin{equation}
\hat{\bphi}(\br) =\left(
\hat{\bmu}^T(\br) , \hat{\bmu}^\dagger(\br)
\right)^T.
\end{equation}
Then up to a constant independent of $\theta$ we find that $ {\cal
H}_{\rm Bog} = \frac{1}{2} \int d \br \hat{\bphi}^\dagger(\br)
{\bf M}(\br,\bp) \hat{\bphi}(\br) $, where
\begin{align}
{\bf M}(\br,\bp) =&\id \otimes \id \,\frac{p^2}{2m}+
\frac{g \rho_0}{2}(\id + \bsx)\otimes (\id +
\bsz)
\\
&-\frac{Q p_y}{m} \left( \id \otimes \bsy \cos(2Qx)-
  \bsz \otimes \bs_\theta \sin(2Qx)
\right).
\notag
\end{align}
In this expression, all $\theta$-dependence is included in $\bs_\theta\equiv
\cos(\theta) \bsx + \sin(\theta) \bsz$ and $\otimes$ is the Kronecker product. This Hamiltonian can be
diagonalized using a symplectic transformation \cite{blaizot86,
powell10}, which amounts to solving the Bogoliubov-de Gennes equations,
\begin{equation}
\label{Eq:Eval} \bet {\bf M}(\br , \bp) {\bf v}_{\bk n}(\br) = \ce_{\bk
n} {\bf v}_{\bk n}(\br),
\end{equation}
for positive eigenvalues $\ce_{\bk n}$, where $\bet=\bsz \otimes
\id$, and  ${\bf v}_{\bk n}(\br)$ is a four-component function.
Because of the translational symmetries of ${\bf M}(\br,\bp)$, the
eigenvalues are labelled with band index $n$ and momentum $\bk$ in
the Brillouin zone (BZ) defined as $-\infty <k_y,k_z<\infty$ and
$-Q\le k_x < Q$. As usual,  the eigenvectors are normalized as
$\bra{\bv_{\bk n} } \bet \ket{\bv_{\bk' n'}} \equiv \int d\br
\bv_{\bk
  n}^\dagger(\br) \bet \bv_{\bk' n'}(\br)
=\delta_{\bk \bk'}\delta_{n n'}  {\rm sgn}(\ce_{\bk n})$.  In
practice, Eq.~(\ref{Eq:Eval})  is simplest to solve in momentum
space.  Since the momentum space representation of ${\bf M}(\br, \bp)$ is
an infinite matrix, in numerical calculations it must be
truncated at high momentum and the eigenvalues of interest must be
checked to be independent of the cutoff.

\begin{figure}
\includegraphics[width=3.5in]{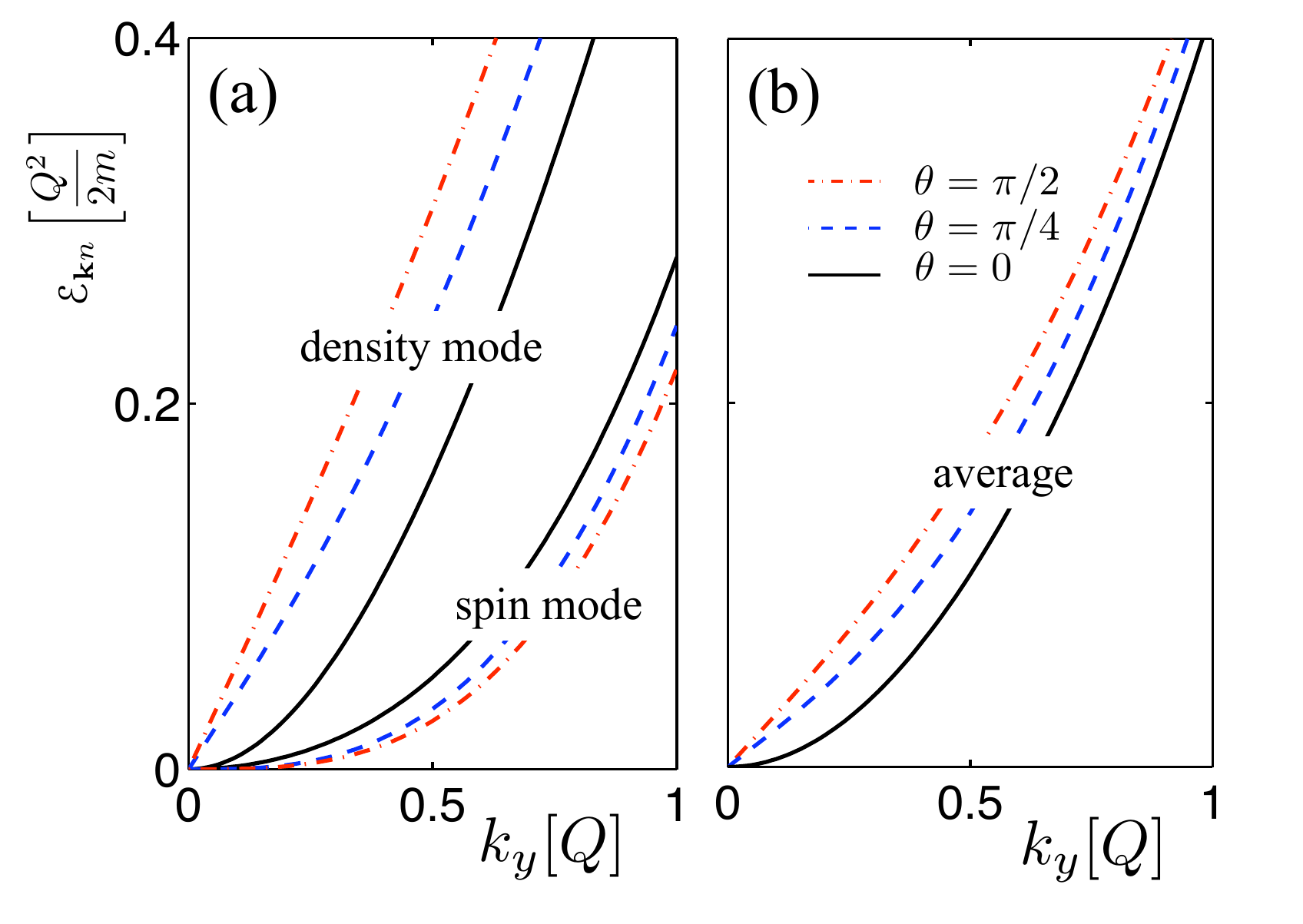}
\caption{(Color online) (a) The dispersions for the density and spin Goldstone modes for three
  values of $\theta$ for $k_x=k_z=0$.  (b) The average (arithmetic
  mean) of the density and spin
  modes. In both plots we have  fixed $g\rho_0=\frac{Q^2}{2m}$.}
\label{Fig:disp}
\end{figure}

In Fig.~\ref{Fig:disp} we show the two gapless (Goldstone) modes
for several values of $\theta$, found numerically from
Eq.~(\ref{Eq:Eval}).  In experiments of \cite{lin11},
$\varepsilon_Q=\frac{Q^2}{2m}\simeq g \rho_0$, so we set these
quantities to be equal.  The dispersion is plotted along $k_y$
since, as can be seen from Eq.~(\ref{Eq:Eval}), the spectrum
$\ce_{\bk n}$ has no $\theta$-dependence when $k_y=0$. We refer to
the dispersions as `density' and `spin' modes since they reduce to
the known expressions $\ce_{\bk
d}=\sqrt{\varepsilon_\bk(\varepsilon_\bk+2g\rho_0)}$ and ${\cal
E}_{\bk s}=\varepsilon_\bk$ in the limiting case of $Q=0$, where
$\varepsilon_\bk=\frac{k^2}{2m}$ is the free particle dispersion.  One
sees that upon increasing $\theta$ from zero to $\pi/2$, the spin mode
decreases in energy while the density mode increases.  This gives, in
a sense, a competing effect in terms of which $(a,b)$ configuration is
selected from fluctuations. Noting this, in the right panel we plot
the average of the spin and density modes for each value of $\theta$.
One sees that the average is always lowest in energy for $\theta=0$.
This indicates that the zero-point fluctuations from the Goldstone modes will select
$\theta=0$ state though things become more subtle for $T>0$.  Such
a state, as can be seen from Eq.~(\ref{Eq:gen2}), corresponds to all
bosons condensing into a single momentum state of the RBEC system.
The order-by-disorder mechanism will be considered more quantitatively below.

Analytical expressions for the dispersions and eigenvectors of
Eq.~(\ref{Eq:Eval}) can be found perturbatively in the
long-wavelength limit $\varepsilon_{\bk} \ll \varepsilon_Q,
g\rho_0$.  
In this limit one finds $ \ce_{\bk d} = \sqrt{ 2 g
\rho_0\left(\varepsilon_{k_{xz}} + \lambda \varepsilon_{k_y}
\sin^2(\theta )\right)} $ and $ \ce_{\bk s} =
\sqrt{\frac{\varepsilon_{k_{xz}}(\varepsilon_{k_{xz}}+\lambda\varepsilon_{k_y})^2}{\varepsilon_{k_{xz}}+\lambda
    \varepsilon_{k_y}\sin^2(\theta)}}
$ for the density and spin modes respectively where
$\lambda=g\rho_0/(4 \varepsilon_Q + 2g\rho_0)$ and $k_{xz}=\sqrt{k_x^2+k_z^2}$.  These 
agree well with the numerical results shown in Fig.~\ref{Fig:disp}
for small $\bk$ except for two special cases which require more
careful analysis. In particular, for $\theta=0$,
the density mode disperses quadratically along $k_y$ while
for $0<\theta\le\pi/2$ the spin mode disperses as $k_y^3$
along $k_y$.  Otherwise the density and spin mode have
respectively linear and quadratic dispersions about their minima.
It is interesting to compare these to the noninteracting energies
shown in Fig.~\ref{Fig:nispec}, which have quadratic and quartic
dispersions about their minima.

\section{Quantum and Thermal Order by Disorder}

\begin{figure}
\includegraphics[width=3.4in]{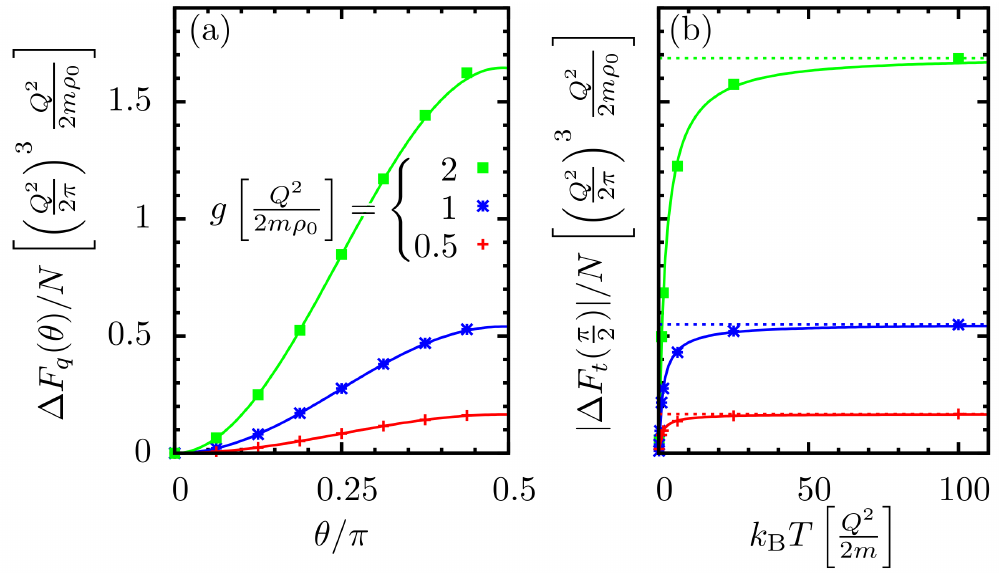}
\caption{(Color online) (a) The zero-point contribution to
the free energy $\Delta F_q$ as a function of $\theta$ for three
different values of $g$.  (b) The absolute value of the (negative)
thermal free energy splitting between the $\theta=0$ and
$\theta=\pi/2$ configurations $|\Delta F_t(\pi/2)|$ as a function of
temperature (solid line).  This is seen to
approach the quantum zero-point splitting $\Delta F_q(\pi/2)$ at high
temperatures (dashed line).  In both panels the solid lines are
fits to the numerically computed data points. }
\label{Fig:combine}
\end{figure}

Let us now consider the free energy due to quantum and thermal
fluctuations described by ${\cal H}_{\rm Bog}$. It is useful to
separate out the contribution from zero-point fluctuations and
write $\cf(\theta) = \cf_q(\theta) + \cf_t(\theta) $ where
\begin{equation}
\cf_q(\theta)= \frac{1}{2} \sum_{\bk \in {\rm BZ} ,n} \ce_{\bk n}(\theta),
\end{equation}
\begin{equation}
\cf_t(\theta)= k_B T \sum_{\bk \in {\rm BZ},n} \ln \left( 1- e^{-\beta  \ce_{\bk n}(\theta)}\right),
\end{equation}
and $\beta=1/k_BT$ is inverse temperature. Reminiscent of the
zero-point photon contribution to the Casimir-Polder force
\cite{casimir48}, the purely quantum contribution $\cf_q(\theta)$
written as it is diverges.  This divergence can be regularized
by subtracting the free energy for a particular mean-field
configuration which we take to have $\theta=0$: $\Delta
\cf_q(\theta) = \cf_q (\theta) - \cf_q (0)$.  This regularized
expression converges \footnote{ After the $n$-summation is
performed, the summand has the asymptotic form proportional to
$k_y^2/k^5$ for large $\bk$ as can be determined perturbatively},
and no renormalization of the effective range of interactions is
needed. The zero-point contribution to the free energy
numerically computed as a function of $\theta$ is shown in
Fig.~\ref{Fig:combine}(a) where the summation is performed over 26
bands  (we emphasize that in order to obtain quantitatively
correct results, including only the gapless modes is
insufficient).  One sees, indeed,  that the $\theta=0$ state has
the lowest energy and at $T=0$ such a state is unambiguously
selected.

We now turn to  the finite-temperature contribution to the
free energy. Interestingly, one finds that the sign of the thermal
contribution $\Delta F_{t}(\theta) = F_{t}(\theta)- F_{t}(0)$ is negative and opposite to
that of $\Delta F_q(\theta)$.  Furthermore, the magnitude of the
thermal contribution is always smaller than the contribution from
zero-point fluctuations, in contrast to typical situations where
thermal fluctuations enhance the degeneracy lifting and are larger
in magnitude for modest temperatures (see, e.g. \cite{turner07}).
Another instance of where quantum and thermal fluctuations select
different states is in Ref.~\cite{toth10}.
The sign of $\Delta F_t$ at low $T$ can be understood by noting
that the spin mode has the lowest energy for $\theta=\pi/2$
(Fig.~\ref{Fig:disp}).

As seen in Fig.~\ref{Fig:combine}(b), the magnitude of $\Delta
F_{t}(\theta)$ approaches  $\Delta F_{q}(\theta)$ at high $T$, so
that $\Delta F (\theta)=\Delta F_q(\theta)+\Delta F_t(\theta)= \mathcal{O}\left(T^{-1}\right) \rightarrow 0$ in this limit.  This
behavior can be understood through a high $T$ expansion of the
free energy
\begin{equation}
F_t(\theta) \approx k_B T  \sum_{\bk \in {\rm BZ} ,n} \ln\left( \beta \ce_{\bk n}(\theta) \right)
-\frac{1}{2} \sum_{\bk \in {\rm BZ} ,n} \ce_{\bk n}(\theta).
\end{equation}
As the second term cancels the quantum contribution,
we focus on the larger first term which can be written as
\begin{align}
\notag
k_B T  \sum_{\bk \in {\rm BZ} ,n} \ln\left( \beta \ce_{\bk n}(\theta)
\right)
&= \frac{1}{2} k_B T  \ln| {\rm det} (\beta \bet {\bf M}(\br, \bp)|\\
\notag
&=k_B T  \sum_{\bk \in {\rm BZ'},n} \ln\left|\beta \lambda_{\bk n}\right|
\end{align}
where $\lambda_{\bk n}$ are the eigenvalues of ${\bf M}(\br,\bp)$ and we
have used $|{\rm det}(\bet)|=1$.   The second summation  above is over
the reduced Brillouin zone ${\rm BZ'}$ which is restricted to
positive values of $k_x$. The eigenvalues $\lambda_{\bk n}$ are
independent of the condensate configuration given by $\theta$.
This can be seen by noting that the
$\theta$-dependence of ${\bf M}(\br,\bp)$ can be removed through the unitary
transformation ${\bf M}(\br,\bp)\rightarrow {\bf U}^\dagger {\bf
  M}(\br,\bp) {\bf U} $
where $ {\bf U}=\frac{1}{2}(\id+\bsx) \otimes \id + \frac{1}{2}(\id -
\bsx) \otimes e^{i\theta \bsy} $ 
\footnote{The same transformation introduces $\theta$-dependence into
  $\bet$ and so the eigenvalues of $\bet {\bf M}(\br,\bp)$ generally depend
  on $\theta$, which determine the Bogliubov spectrum. }.      
Thus to this order we find that
$\Delta F_t(\theta) = - \Delta F_q(\theta)$.  The next-order term
in the high-temperature expansion has $1/T$ dependence which is
evident in the numerical results shown in
Fig.~\ref{Fig:combine}(b).

\section{Condensate Depletion}

Having established using the Bogoliubov expansion that fluctuations
select $\theta=0$, we now investigate the self-consistency of this
approach. This is determined by the depletion or the number of
particles excited out of the condensate $N_{\rm ex}$ considered as a
fraction of the total particle number $N$. Consistency of course
requires that this be finite, while neglecting of terms beyond
quadratic order in $\cal{H}_{\rm Bog}$ is quantitatively reliable only if
$N_{\rm ex} \ll N$. The quantum and thermal contributions to $N_{\rm
ex} = N_q + N_t$ are, respectively, $N_q=\frac{1}{2}\sum_{\bk \in {\rm
BZ}, n} \bra{v_{\bk n}}(\id - \bet)\ket{ v_{\bk n}}$ and
$N_t=\sum_{\bk \in {\rm BZ}, n} \bra{v_{\bk n}}\ketn{v_{\bk n}}
f(\ce_{\bk n})$, where $f(x)=(e^{\beta x}-1)^{-1}$ is the
Bose-Einstein distribution function. The only possible divergences of
these expressions are in the infrared, and so can be studied
analytically using the small-$k$ expansion.

In 3D at $T = 0$, $N_q$ is finite and so can be sufficiently small
provided weak-enough interactions. Our numerical results 
in fact demonstrate that the depletion is small even
for moderately strong interactions, including in the region of
experimental relevance. For $T>0$, the thermal contribution is instead
found to have a logarithmic divergence in 3D. This divergence (similar
to that occurring in quasi-2D scalar condensates) is naturally removed
for finite-sized systems and the condensate will thereby satisfy the
stability criterion. For 2D condensates with isotropic SO coupling,
the situation is different. Here, the quantum depletion again
converges, but the thermal depletion diverges as $1/k$ for small
$k$. Thus, at $T>0$ our theory is unstable in 2D, which is consistent
with work suggesting fragmentation \cite{stanescu08,gopalakrishnan11}.
Our conclusions on the stability of the condensate
are, remarkably, identical to those based on the simple application of
the Einstein criterion to the noninteracting spectrum. This is
particularly surprising since the low-$k$ region is strongly modified
by the interactions, giving quasiparticle modes that disperse with
different powers than in the noninteracting case.

\section{Experimental Feasibility}

We now comment on the experimental
feasibility of observing order by disorder in RBEC.  
We first consider the magnitude of the degeneracy lifting.  As a prototypical
example we take spin one  $^{87}$Rb.  For a typical density of $\rho_0=2 \times
10^{14}{\rm cm}^{-3}$ and $g\rho_0=\frac{Q^2}{2m}$, we find (for the
appropriate scattering lengths) that at
zero temperature the free energy splitting per particle due to fluctuations is
$\Delta F(\pi/2)/k_B N= 110 \, {\rm pK}$.  One should note that this number
should not be directly compared with the condensate temperature since
the \emph{total} energy determines the ground state.  It is this
energy which will determine experimental timescales for the relaxation
to the ground state.
Spin-one
atoms also possess a spin-dependent interaction term which
we have neglected. For $^{87}$Rb, however, this spin-dependent
interaction is relatively small (.5\% that of the spin-independent)
and as a result the degeneracy lifting from fluctuations is typically
larger.  Alternatively, one could use fermionic homonuclear molecules
that have a singlet ground state.  More importantly, schemes to create
SO coupling in bosonic systems typically rely on utilizing
dressed states \cite{dalibard10, lin11, zhang11} which can induce anisotropic
interactions.  The magnitude of such terms and their effect on the
order-by-disorder mechanism will be left to future work when it becomes
clear which of the several proposed schemes is most promising to
realize Rashba coupling.

Another entity of experimental relevance is the harmonic confining
potential.  The results of the current manuscript will hold if the
conditions for the local density approximation (LDA) are satisfied
\cite{dalfovo99}.  
For our system this requires that the energy splitting of  
the single-particle states  as recently found in  
in \cite{hu11,sinha11} be small compared to the interaction energy. This energy  
splitting becomes small for weak trapping and/or strong SO coupling,  
resulting in a large quasi-degenerate manifold of single-particle  states.
Conversely, in the weakly interacting limit (where
LDA is inapplicable) recent work \cite{hu11,sinha11} has shown the
ground states of the RBEC system in a trap can form vortex lattices.

\section{Conclusion}

In conclusion we have investigated the system of Rashba SO-coupled
bosons with isotropic interactions (RBEC).  In general bosons with SO
coupling offer a genuinely new class of systems which has not been addressed
in the vast solid state literature on spintronics \cite{zutic04}.
In particular, we have established that fluctuations select the RBEC
system to condense into a single momentum state.  We have argued that
such a configuration is stable in 3D but destabilized when $T>0$
in 2D.  We expect bosons with Rashba SO coupling to be
realized in the near future for which the predicted configuration
will be observable in Stern-Gerlach experiments. In future
studies it will be interesting to investigate more general
combinations of Rashba and Dresselhaus SO coupling.  Such systems also
possess accidental mean-field degeneracies and thus fluctuations are
expected to play an important role in determining their ground states.
In addition it will be interesting to investigate RBEC systems in two 
dimensions.

\acknowledgements

We thank Ian Spielman and Anna Maria Rey for discussions and J. T. 
Chalker for drawing our attention to Ref.~\cite{chalker11}.  This
work was supported by JQI-NSF-PFC, JQI-ARO-MURI, ERC Grant QUAGATUA,
Spanish MINCIN grant TOQATA, and  KITP under grant NSF PHY05-51164,
where this work was initiated.


%

\end{document}